# Arbitrarily routed mode-division multiplexed photonic circuits for dense integration

Yingjie Liu[1], Ke Xu[1,*], Shuai Wang1, Weihong Shen[3], Hucheng Xie[1], Yujie Wang[1], Shumin Xiao[1, 2,3], Yong Yao[1,*], Jiangbing Du[3,*], Zuyuan He[3], and Qinghai Song[1, 2,*].

1.  State Key Laboratory on Tunable laser Technology, Ministry of Industry and Information Technology Key Lab of Micro-Nano Optoelectronic Information System, Harbin Institute of Technology (Shenzhen), Shenzhen, P. R. China, 518055.
2.  Collaborative Innovation Center of Extreme Optics, Shanxi University, Taiyuan 030006 P. R. China.
3.  State Key Laboratory of Advanced Optical Communication Systems and Networks, Shanghai Jiao Tong University, Shanghai 200240, China.

**ABSTRACT**

Mode-division multiplexing (MDM) is becoming an enabling technique for large-capacity data communications via encoding the information on orthogonal guiding modes. However, the on-chip routing of a multimode waveguide occupies too large chip area due to the constraints on inter-mode cross talk and mode leakage. Very recently, many efforts have been made to shrink the footprint of individual element like bending and crossing, but the devices still occupy >10×10 μm$^2$ footprint for three-mode multiplexed signals and the high-speed signal transmission has not been demonstrated yet. In this work, we demonstrate the first MDM circuits based on digitized meta-structures which have extremely compact footprints. The radius for a three-mode bending is only 3.9 μm and the footprint of a crossing is only 8×8 μm$^2$. The 3×100 Gbit/s mode-multiplexed signals are arbitrarily routed through the circuits consists of many sharp bends and compact crossing with a bit error rate under forward error correction limit. This work is a significant step towards the large-scale and dense integration of MDM photonic integrated circuits.

**Introduction**

The orthogonal spatial modes in integrated optical waveguides are considered as an important degree of freedom to boost the data capacity for optical interconnect[1-4]. For large-capacity optical interconnect, mode-division multiplexing (MDM) well serves as an alternative technique to the wavelength division multiplexing since it only needs monochromatic laser and precise wavelength control is not necessary[5]. To realize a functional on-chip MDM circuit[6-8], various multimode devices are required such as (de)multiplexer[9-11], grating coupler[12], switch[13-15], mode filter[16], splitter[17,18] and many other building blocks[19-21]. More importantly, the interconnections among these elements have to rely on multimode waveguides with a few orthogonal and co-propagating modes. This is quite challenging, as the transmission of a MDM signal in such multimode photonic circuits is vulnerable to sharp bending and cross connection due to the radiation leakage and inter-mode coupling. The adiabatic wave propagation requires too large footprint for a bent waveguide[22]. The mode-independent crossing is also complicated for higher order modes since the MDM signals needs to be demultiplexed to single modes prior to waveguide crossing[23]. For both cases, the MDM signal is routed at the cost of large chip area which eventually limits the integration density.

To arbitrarily route the multimode waveguide, the bending and crossing are two key elements. Recently, various waveguide structures have been proposed to achieve low-loss and low inter-mode-crosstalk waveguiding for the bend and cross connection scenario[24-29]. However, most of the demonstrations only achieve dual-mode bending and crossing. L. H. Gabrielli et al have shown a three-mode bending with low inter-mode coupling, but the bending raidus is large and grey-scale lithography is used. A three-mode crossing has been recently reported based on Maxwell fish eye lens structure with a footprint of 11.6 ×11.6 μm$^2$ where a precisely designed focusing taper is needed.

In this article, we demonstrate the first arbitrarily routed circuit for three-mode-multiplexed signals via ultra-compact bending (2.75-μm curvature radius) and crossing (8×8 μm$^2$). The sharp bending and direct cross connection for TE0, TE1, and TE2 modes are realized by the intriguing wave guiding enabled by a discretized meta-structure with a number of nanoholes. The position of the nanoholes are determined by optimization algorithm, which enables local index engineering within the device region. The tailored index profile allows for mode matching between the multimode waveguide and the bending/crossing region. Furthermore, our proposed devices are fully compatible with the complementary metal-oxide-semiconductor (CMOS) fabrication process. We have achieved a three-mode bending with 0.7-0.9 dB loss and a crossing with 0.3-0.7 dB loss over 80 nm bandwidth. Finally, three-mode division multiplexed signal with 3×100 Gbit/s data rate is successfully routed along the arbitrarily designed circuits with a bit error rate (BER) under forward error correction (FEC) limit.

**Results**

**Three-mode bending:** The micro-bend with a digital meta-structure is designed and illustrated in Figure 1(a). The bending region is divided into discrete pixels with a minimum feature size of 130 nm. The device platform is silicon-on-insulator (SOI) with 220nm top silicon and 2 μm buried oxide. We define a binary material state for each pixel, i.e. it can be either unetched silicon or air by etching away the silicon. This allows for a local index contrast of Δn=2.48 to manipulate the wave propagation. The waveguide width is chosen to be 2.3 μm which can support three lowest order modes and allow for enough number of pixels as well. For the first time, we design an unprecedented compact bend with only 2.75 μm inner curvature radius (effective radius: 3.9 μm) for a three-mode division multiplexed signal. The bend enables simultaneous transmission of $TE_0$, $TE_1$ and $TE_2$ mode with low loss and low inter-mode crosstalk. The optical field distributions for the lowest three modes are simulated by 3-dimensional finite difference time domain (FDTD) method and shown in Figure 1 (b)-(d). It can be seen that, the optical waves for all the modes are squeezed and focused to the corner region of the bend by a few nanoholes located near the inner sidewall of the bent waveguide. The outer bound nearly "reflects" the beam to the output waveguide with preserved mode profile. Though this is not an absolute adiabatic process, the inter-mode interference can be well suppressed to a negligible level. The nanohole distributions can be optimized by many algorithms[30-35], and here we use a direct binary search algorithm (see methods). We also quantitatively study how these nanoholes affect the device performance by filling some of the holes with silicon and re-simulate the transmission spectra. We found that the three isolated holes near the outer curvature has nearly negligible impact on the transmission for all the modes. The efficiency drops less than 2% without these air holes. However, the other nanoholes are critical to the device performance (see supplementary information). The simulated spectra of the device from 1500 nm to 1580 nm are shown in Figure 1 (e)-(g). The simulated insertion loss (ILs) are 0.56 dB, 0.76 dB, 0.81 dB for $TE_0$, $TE_1$ and $TE_2$ mode, respectively. The crosstalk (CTs) are lower than -20dB for all the modes over the 80 nm wavelength range.

The devices are fabricated via CMOS compatible silicon photonic process and the details can be found in Methods. To demonstrate the sharp turning of a three-mode multiplexed signal, we fabricated an MDM circuit consisting of a mode multiplexer (MUX), two cascaded bends and a mode demultiplexer (DEMUX), as shown in Figure 2 (a). The MUX is also a compact digital meta-structure with a footprint of only 3.4×3.9 μm² which is described in details in the supplementary information (SI). As illustrated in Figure 2 (a), this is a 3×3 circuit with three input ports I1 ($TE_0$ mode), I2 ($TE_1$ mode), I3 ($TE_2$ mode) and three output ports labeled as O1~O3 ($TE_0$-$TE_2$ modes), respectively. In the meantime, as shown in Figure 2 (b), an extra reference circuit with a back-to-back mode MUX is fabricated to normalize the transmission of the device and to extract the bending loss. The TE-polarized grating coupler is used to interface the fiber and the waveguide, which has around 6 dB loss per grating. The meta-structure-based bent waveguide is well fabricated as we designed which can be seen from the zoom-in scan electron microscope (SEM) image as shown in Figure 2 (c). The measurement setup for the transmission spectra of the device are described in the methods. The normalized transmission spectra measured at O1~O3 ports are plotted in Fig. 2 (d)-(f) when the continuous wave is launched from input ports 11~13, respectively. The bend is measured to have ILs of 0.78 dB, 0.8 dB, 0.9 dB for $TE_0$, $TE_1$ and $TE_2$ mode, respectively. The measured CTs are lower than -20dB for all the guiding modes from 1500-1580 nm which is quite consistent with the simulation results.

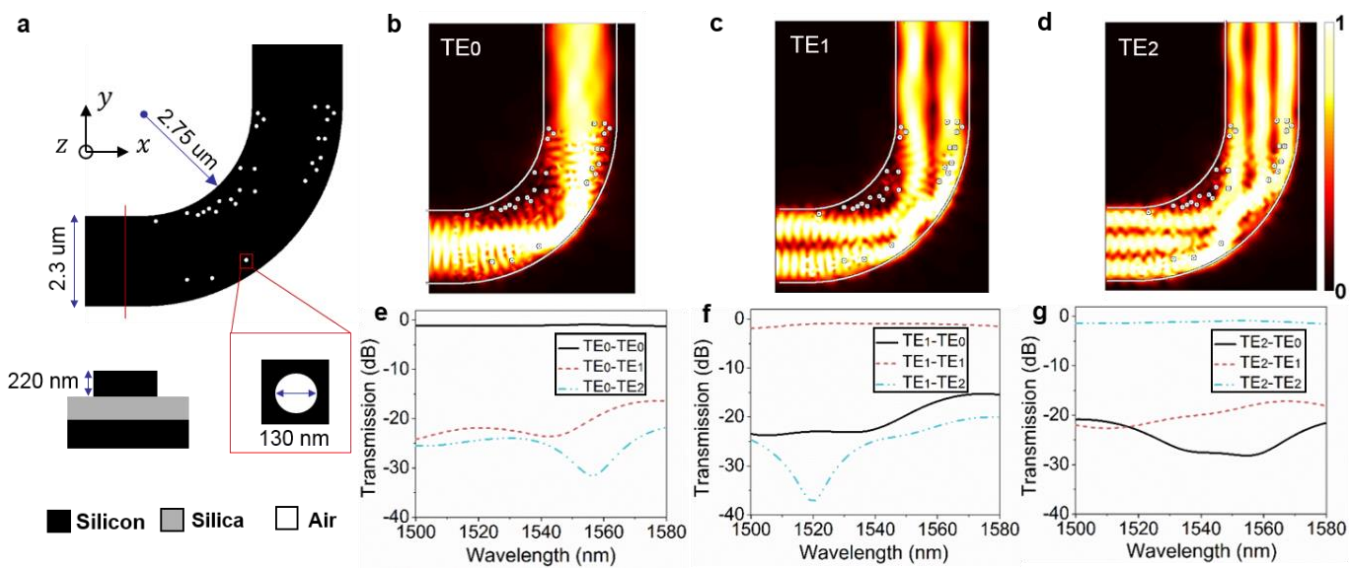

**Figure 1 | Design and calculation of the digital micro-bend.** (**a**) Schematic of the designed 3-mode bending with a digital meta-structure. (**b**) - (**d**) The simulated optical field distribution for micro-bend with three modes ($TE_0$-$TE_2$). (**e**) - (**g**) The simulated transmission spectra for micro-bend with three modes ($TE_0$-$TE_2$).

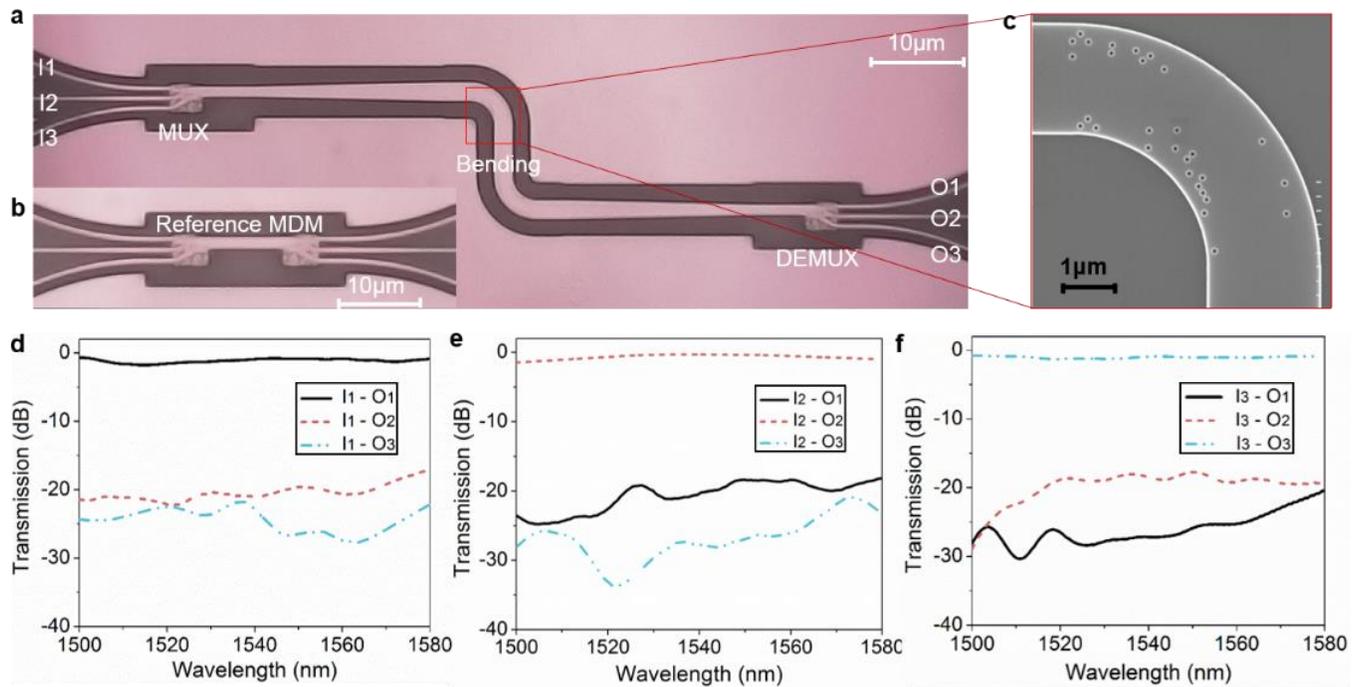

**Figure 2 | Fabrication and experimental demonstration of the digital micro-bend.** (**a**) The top-view microscope image of the MDM circuit consisting of MUX, bending and DEMUX. (**b**) The microscope image of the reference circuit: a back to back (de)MUX. (**c**) The SEM image of the fabricated bending. (**d**) - (**f**) The measured transmission spectra of the MDM circuit for the optical wave launched from I1 (TE$_0$ mode), I2 (TE$_1$ mode), I3 (TE$_2$ mode), respectively.

**Three-mode crossing:** The on-chip routing inevitably needs to address the waveguide cross-connect problem. Here we design a three-mode crossing with a footprint of 8×8 μm², and the structural schematic is described in Figure 3 (a). The device has a quadrature symmetry with the minimum feature size of 130 nm. The input/output waveguide widths of the crossing are set to 1.4 μm to support TE$_0$, TE$_1$ and TE$_2$ modes. Each quadrature area is axial symmetric and occupies only 4×4 μm². This design window is divided into 25×25 discrete pixels (fully etched holes). The optimization process can also be referred to the Methods. Fig. 3 (b) - (d) show the simulated in-plane optical field profiles of the nanostructured crossing when the input waves are TE$_0$, TE$_1$ and TE$_2$ mode, respectively. There are several nanoholes located at the corner region of the crossing, which provide a forbidden gap with low index to avoid mode expansion. Hence, the mode profile of each mode can be well preserved when propagating through the crossing region. There is a small amount of mode leakage into the orthogonal waveguide. This can be further reduced by using a power monitor closer to the crossing and set an additional optimization objective to reduce such leaked power. To quantitively investigate the performance of the crossing, we calculate the transmission efficiency and the mode crosstalk considering mode overlap integral in the simulations. Fig. 3 (e) - (g) show the calculated transmission and CT spectra. The simulated ILs for TE$_0$, TE$_1$ TE$_2$ mode are 0.22 dB, 0.55 dB, and 0.6 dB, respectively. The simulated CTs are lower than -30 dB for all modes from 1500 nm to 1580nm wavelength range.

We experimentally characterize the crossing performance by fabricating the 3-mode cross-connect structure, and the microscope image of the circuit is shown in Figure 4 (a). The devices were fabricated using the same process with the bending. Due to the ultra-low loss of an individual crossing, we cascade different numbers of crossing to accurately obtain the IL of the device. Figure 4 (b) is the microscope image of a 4-stage cascaded crossing circuit which is more measureable than a single crossing. The zoom-in SEM image of the nanostructured crossing is shown in Figure 4 (c) which confirms that the desired pattern has been successfully transferred into the device layer. The transmission spectra of the cascaded crossing are measured and normalized. The measured ILs and CTs for TE$_0$ - TE$_2$ modes transmitted through a single crossing is shown in Figure 4 (d) – (f), respectively. The ILs are 0.3 dB, 0.56 dB, 0.7 dB for TE$_0$, TE$_1$ and TE$_2$ mode, respectively. The CTs for all modes are characterized by measuring the optical power at the other output ports. The spectra indicates that the CTs are of the orders of -30 dB in average for all the three guiding modes.

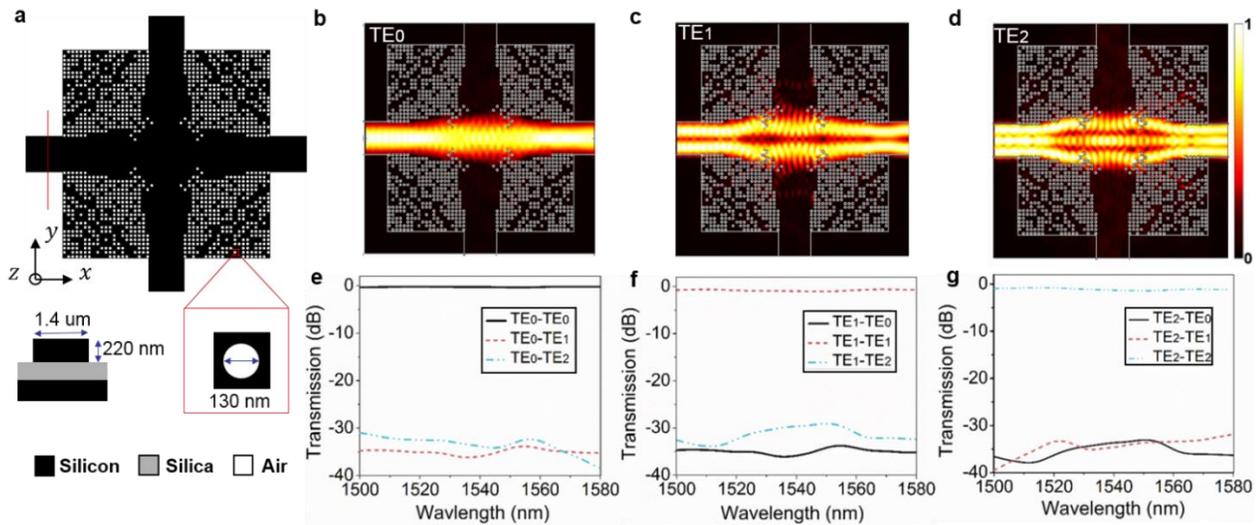

**Figure 3 | Design and calculation of three-mode crossing.** (**a**) Schematic of the designed 3-mode crossing with a digital meta-structure. (**b**) - (**d**) The simulated optical field distribution for 3-mode crossing with three modes (TE$_0$-TE$_2$). (**e**) - (**g**) The simulated transmission spectra for micro-bend with three modes (TE$_0$-TE$_2$).

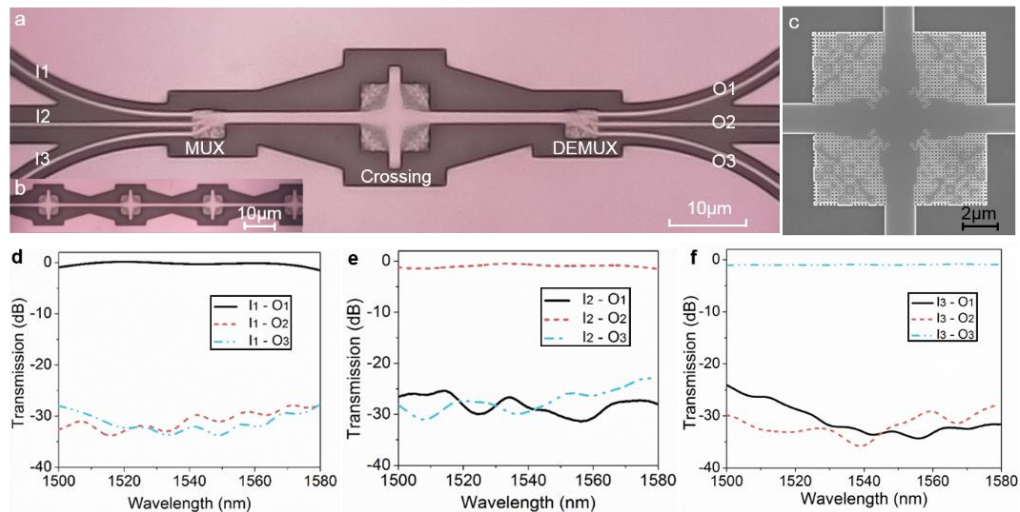

**Figure 4 | Fabrication and experimental demonstration of three-mode crossing.** (**a**) The top-view microscope image of the MDM circuit consisting of MUX, crossing and DEMUX. (**b**) The microscope image of 4 cascaded crossing. (**c**) The SEM image of the fabricated crossing. (**c**) - (**e**) The measured transmission spectra for the MDM circuit.

**MDM routing circuits:** Most of the previous demonstrations on MDM circuit hardly routed the multimode waveguides due to the limited capability of sharp turning and compact cross connection. Here, we verify the possibility of arbitrary and compact on-chip routing of a 3×100 Gbit/s MDM signals via multimode waveguide without the need of demultiplexing. In principle, the MDM signal can be delivered to any chip locations via any routing path provided the low loss bending and crossing are available. For a proof-of-concept demonstration, we arbitrarily design two MDM circuits with the proposed bending and crossing. MDM circuit1 consists of a MUX, 4 bends, a crossing, and a DEMUX. The microscope image of the three-mode division multiplexed circuit is shown in Figure 5 (a). Such circuit allows both sharp bending and cross connection which can be extended to cascaded structure or waveguide array scenarios. We define input ports I1-I3 and output ports O1-O3 for TE$_0$ - TE$_2$ modes, respectively. Figure 5 (b) and (c) show the measured ILs and CTs spectra for all the modes after transmission through the circuit. The losses including the IL of MUX, DEMUX, two times cross connect and four bending. The average ILs are ~ 8 dB and the CTs are lower than -25dB within the wavelength range from 1500 - 1580 nm.

MDM circuit2 consists of a MUX, 10 bends, and a DEMUX. The waveguide is configured in a spiral route and can be further scaled to a dense spiral waveguide. Such a circuit is used to confirm the ability to change the waveguide routing direction arbitrarily. The microscope image of MDM circuit2 is shown in Figure 6 (a). The measured ILs and CTs spectra for all the modes after transmission through the circuit are shown in Figure 6 (b)-(c). The average ILs for all the mode channels are 13 dB from 1500-1580 nm, and the CTs of MDM circuit2 is measured to be lower than 30 dB within the same wavelength range.

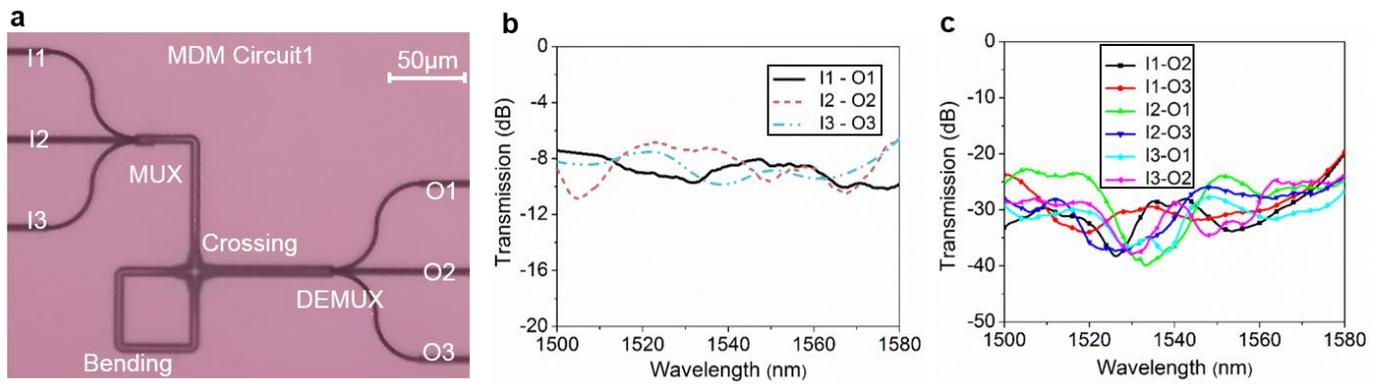

**Figure 5 | Fabrication and experimental demonstration of the MDM circuit1.** (a) The top-view microscope image of the MDM circuit1 consists of a MUX, 4 bends, a crossing, and a DEMUX. (b) The measured ILs for different input/output ports. (b) The measured CTs for different input/output ports.

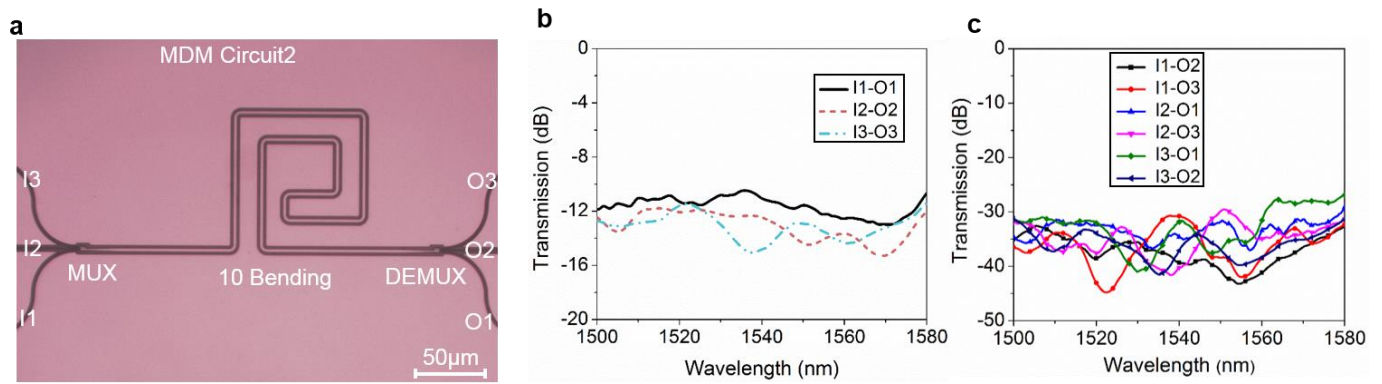

**Figure 6 | Fabrication and experimental demonstration of the MDM circuit2.** (a) The top-view microscope image of the MDM circuit2 consists of a MUX, 10 bends and a DEMUX. (b) The measured ILs for different input/output ports. (b) The measured CTs for different input/output ports.

In addition to simply measuring the circuit loss and cross talk, we verify the capability of 3×100 Gbit/s MDM signal routing via the proposed circuits. We demonstrate a 3-channel MDM on-chip routing circuits with each guiding mode encoded by a 100 Gbit/s discrete multi-tone (DMT) signal. The detail experimental setup and the digital signal processing technique can be found in the supplement information. For each channel, we have successfully routed the signal through circuit1 and circuit2. The output signal is direct detected with the signal to noise ratios plotted in Figure 7. It can be seen from the SNR curve that the multimode circuits induce very slight degradation to the signal quality. The maximum bit loading is 5 which corresponds to the 32-QAM format. The spectrally efficient DMT technique allows us to achieve 100Gbit/s single lane rate under a bandwidth constrained condition. The constellations with a bit index of 5 (32-QAM) and 4 (16-QAM) under different subcarrier bands are plotted as the inset which indicate a good signal quality for high order modulation formats. The bit error rate (BER) measurement is performed and the results can be found in the supplement information. The BER for transmission in both circuits are well below the forward error correction (20% overhead) limit. Thus, the ultra-compact signal routing of a 3×100 Gbit/s MDM signal is successfully achieved.

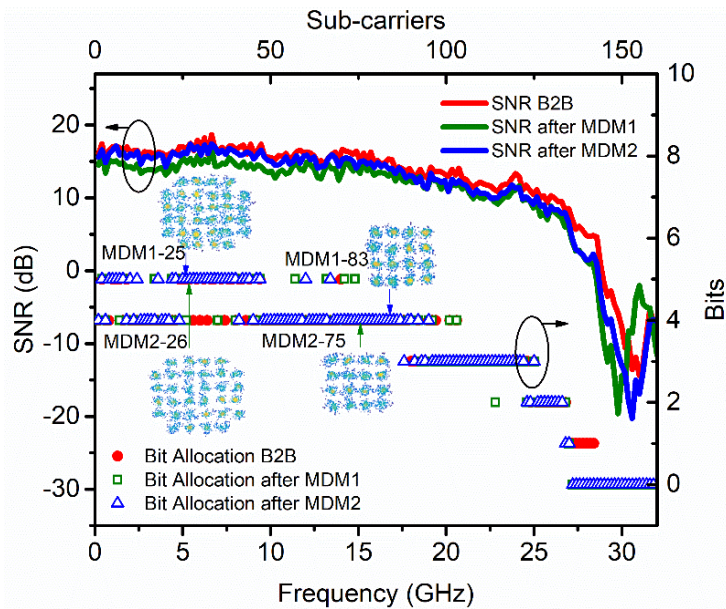

**Figure 7 | The bit loading curve.** The SNR under different frequencies and the bit loading for different subcarrier. Inset: constellations of different frequency band.

**Discussion**

For the first time, we demonstrate the MDM circuit which can be arbitrarily routed using extremely compact bending and crossing. Manipulating the optical wave in such a tiny structure relies on the index engineering in a deep subwavelength scale. An optimized index profile is the key to mode matching which helps us to avoid mode leakage and inter-mode coupling. The index contrast of the nanoholes with air cladding is $\Delta n=2.48$. However, a thick oxide cladding is normally deposited on the silicon layer which reduces the index contrast by 20%. Here, we have confirmed in the simulations that the oxide cladding does not affect the design of such compact MDM device. The digitized meta-structure for bending and crossing with oxide cladding are designed with comparable performance with air cladding devices (see supplement information). It is also important to note that even though the device is fabricated via E-beam lithography, the minimum feature of 130 nm can be easily patterned in a 90 nm CMOS fabrication process.

It is important to analyze the contributions made from those discretized nanoholes. We have numerically investigated the performance of the bending and crossing structure if all the etched pixels are filled by silicon again. For such a sharp bend and compact crossing waveguide, the ILs and CTs increases significantly as we expect (See the supplement information). This in turn proves the evidence of the effectiveness of index engineering via inverse design.

**Methods**

**Direct binary search (DBS) method:**
DBS method is easy to implement and fast to convergence which is very suitable for the optimization of digital meta-structure with binary material state. The design area of the device is discretized into pixels. The shape of the pixel can be either circular or square, and the minimum feature size can be determined according to the fabrication capability. Here, each pixel has a binary state of the material property: silicon or air. The iterative optimization process began with an all silicon initial structure. Then we set the figure-of-merits (FOMs), convergence condition and objectives. FOMs are numerically calculated by 3D FDTD simulation with 30nm×30nm×30nm grid size. The program iteratively searches the pixel distribution towards the design target until the convergence condition is met. The convergent results are obtained after 4 iterations for each device in this work. The whole design process is performed via an 8-core server. For the bending device, it takes ~ 50 hours in average to get the convergent results after 4 iterations**.** For the crossing device, it takes ~ 40 hours in average to get the convergent results after 3 iterations**.**

**Device fabrication process:**
The devices are fabricated on a SOI wafer with 220 nm top silicon device layer on 2-μm buried oxide. The positive photo resist ZEP 520A is used as the soft mask on SOI. The devices are patterned by the electron beam lithography (Raith eLINE) which operates at 30 kV, then the top silicon layer was fully etched to a depth of 220 nm by using a single-step inductively coupled plasma (ICP) dry etching.

**Measurement experimental setup and methods:**

For device characterization, the setup consists of a 1550nm tunable laser (Yenista T100S-HP/SCL), a fiber-chip coupling stage and a benchtop power meter. The silicon waveguides are accessed by single mode fibers and grating couplers. The tunable laser output is aligned to TE polarization using a polarization controller and then is coupled onto the chip through a pair of grating couplers designed for TE polarization. The grating coupler is designed to be 10-degree tilt with respect to the chip normal direction to suppress the back-reflection. The coupling efficiency of each grating coupler is measured to be ~ 6dB. A benchtop power meter is used to measure the output power of the chip. For high-speed experiments, the setup is described in SI.

**Reference**
1. Dai, D., Wang, J., & Shi, Y. Silicon mode (de)multiplexer enabling high capacity photonic networks-on-chip with a single-wavelength-carrier light. *Opt. Lett*. **38**, 1422-1424. (2013).
2. Wu, X., Huang, C., Xu, K., Zhou, W., Shu, C., & Tsang H. 3X104 Gb/s single-l interconnect of mode-division multiplexed network with a multicore fiber. *J. Lightw. Technol*. **36**, 318-324. (2018).
3. Luo, L., Ophir, N., Chen, C., Gabrielli, L., Poitras, C., Bergmen, K., & Lipson, M. WDM-compatible mode-division multiplexing on a silicon chip. *Nat. Comm*. **5**, 3069. (2014).
4. Hsu, Y., Chuang, C., Wu, X., Chen, G., Hsu, C., Chang, Y., Chow, C., Chen, J., Lai, Y., Yeh, C., & Tsang H. 2.6 Tbit/s on-chip optical interconnect supporting mode-division-multiplexing and PAM-4 signal. *Photon. Technol. Lett*. **30**, 1052-1055. (2018).
5. Dong P. Silicon photonic integrated circuits for wavelength-division multiplexing. *J. Sel. Topics in Quantum Electron*. **22**, 6100609. (2016).
6. Kittlaus, E., Otterstrom, N., & Rakich, P. On-chip inter-modal Brillouin scattering. *Nat. Comm.* **8**, 15819. (2017)
7. Kittlaus, E., Otterstrom, N., Kharel, P., Gertler, S., & Rakich, P. Non-reciprocal interband Brillouin modulation. *Nat. Photon.* **12**, 613-619. (2018)
8. Feng, L., Zhang, M., Zhou, Z., Li, M., Xiong, X., Yu, L., Shi, B., Guo, G., Dai, D., Ren, X., & Guo, G. On-chip coherent conversion of photonic quantum entanglement between different degrees of freedom. *Nat. Comm.* **7**, 11985. (2016).
9. Ding, Y., Xu, J., Ros, F., Huang, B., Ou, H., & Peucheret, C. On-chip two-mode division multiplexing using tapered directional coupler-based mode multiplexer and demultiplxer. *Opt. Express*. **21**, 10376-10382. (2013).
10. Driscoll, J., Grote, R., Souhan, B., Dadap, J., Lu, M., & Osgood, R. Asymmetric Y junctions in silicon waveguides for on-chip mode-division multiplexing. *Opt. Lett*. **11**, 1854-1856. (2013).
11. Wang, J., Chen, P., Chen, S., Shi., Y., & Dai, D. Improved 8-channel silicon mode demultiplexer. *Opt. Express.* **22**, 12799-12807. (2014).
12. Lai, Y., Yu, Y., Fu, S., Xu, J., Shum, P., & Zhang, X. Compact double-part grating coupler for higher-order mode coupling. *Opt. Lett.* **43**, 3172-3175. (2018).
13. Stern, B., Zhu, X., Chen, C., Tzuang, L., Cardenas, J., Bergman, K., & Lipson, M. On-chip mode-division multiplexing switch. *Optica.* **2**, 530-535. (2015).
14. Jia, H., Zhou, T., Zhang, L., Ding, J., Fu, X., & Yang, L. Optical switch compatible with wavelength division multiplexing and mode division multiplexing for photonic networks-on-chip. *Opt. Express.* **25**, 20698-20707. (2017).
15. Xiong, Y., Priti, R., & Ladouceur, O. High-speed two-mode switch for mode-division multiplexing optical networks. *Optica.* **4**, 1098-1102. (2017).
16. Guan, X., Ding, Y., & Frandsen, L. Ultra-compact broadband higher order-mode pass filter fabricated in a silicon waveguide for multimode photonics. *Opt. Lett.* **40**, 3893-3896. (2015).
17. Chang, W., Ren, X., Ao, Y., Lu, L., Cheng, M., Deng, L., Liu, D., & Zhang, M. Inverse design and demonstration of an ultracompact broadband dual-mode 3 dB power splitter. *Opt. Express.* **26**, 24135-24144. (2018).
18. Xu, H., & Shi, Y. Ultra-broadband dual-mode 3 dB power splitter based on a Y-junction assisted with mode converters. *Opt. Lett.* **41**, 5047-5050. (2016).
19. Frandsen, L., Elesin, Y., Frellsen, L., Mitrovic, M., Ding, Y., Sigmund, O., & Yvind, K. Topology optimized mode conversion in a photonic crystal waveguide fabricated in silicon-on-insulator material. *Opt. Express.* **22**, 8525-8532. (2014).
20. Jia, H., Zhou, T., Fu, X., Ding, J., & Yang, L. Inverse-design and demonstration of ultracompact silicon meta-structure mode exchange device. *ACS Photon.* **5**, 1833-1838. (2018).
21. Dorin, B., & Ye, W. Two-mode division multiplexing in a silicon-on-insulator ring resonator. *Opt. Express*. **22**, 4547-4558. (2014).
22. Jiang, X., Wu, H., & Dai, D. Low-loss and low-crosstalk multimode waveguide bend on silicon. *Opt. Express*. **26**, 17680-17689. (2018).
23. Chang, W., Lu, L., Ren, X., Lu, L., Cheng, M., Liu, D., & Zhang, M., An ultracompact multimode waveguide crossing based on subwavelength asymmetric Y-junction. *Photon. J.* **10**, 1-8. (2018).

24. Sun, C., Yu, Y., & Zhang, X., Ultra-compact waveguide crossing for a mode-division multiplexing optical network. *Opt. Lett*. **42**, 4913-4916. (2017).
25. Sun, C., Yu, Y., Chen, G., & Zhang, X., Ultra-compact bent multimode silicon waveguide with ultralow inter-mode crosstalk. *Opt. Lett*. **42**, 3004-3007. (2017).
26. Chang, W., Lu, L., Ren, X., Li, D, Pan, Z., Cheng, M., Liu, D., & Zhang, M., Ultracompact dual-mode waveguide crossing based on subwavelength multimode-interference couplers. *Photon. Res*. **6**, 660-665. (2018).
27. Liu, Y., Sun, W., Xie, H., Zhang, N., Xu, K., Yao, Y., Xiao, S., and Song, Q., Very sharp adiabatic bends based on an inverse design. *Opt. Lett*. **42**, 2482-2485. (2018).
28. Xu, H., & Shi, Y., Metamaterial-based Maxwell's fisheye lens for multimode waveguide crossing. *Laser & Photon. Rev*. **12**, 1800094. (2018).
29. Xu, H., & Shi, Y., Ultra-sharp multi-mode waveguide bending assisted with metamaterial-based mode converters. *Laser & Photon. Rev*. **12**, 1700240. (2018).
30. Shen, B., Polson, R., & Menon, R., Increasing the density of passive photoni-integrated circuits via nanophotonic cloaking. *Nat. Comm*. **7**, 13126. (2016).
31. Piggott, A., Lu, J., Lagoudakis, K., Petykiewicz, J., Babinec, T., & Vuckovic, J., Inverse design and demonstration of a compact and broadband on-chip wavelength demultiplexer. *Nat. Photon*. **9**, 374-378. (2015).
32. Lu, J., & Vuckovic, J., Inverse design of nanophotonic structures using complementary convex optimization. *Opt. Express*. **18**, 3793-3804. (2010).
33. Covey, J., & Chen, R. T., Efficient perfectly vertical fiber-to-chip grating coupler for silicon horizontal multiple slot waveguides. *Opt. Express*. **21**, 10886-10896. (2013).
34. Michaels, A., & Yablonovitch, E., Inverse design of near unity efficiency perfectly vertical grating couplers. *Opt. Express*. **26**, 4766-4779. (2018).
35. Ma, W., Cheng, F., & Liu, Y., Deep-learning-enabled on-demand design of chiral metamaterials. *ACS Nano*. **12**, 6326-6334. (2018).